\documentclass[12pt]{article}
\usepackage{putex}
\usepackage{graphicx}

\begin{document}

\title{Conformal symmetry and the Balitsky-Kovchegov equation}

\authors{Steven S. Gubser}

\institution{PU}{Joseph Henry Laboratories, Princeton University, Princeton, NJ 08544}

\abstract{Solutions to the Balitsky-Kovchegov equation are considered which respect an $SO(3)$ subgroup of the conformal group.  The symmetry dictates a specific dependence of the saturation scale on the impact parameter.  Applications to deep inelastic scattering are considered.}

\preprint{PUPT-2364}

\maketitle
\tableofcontents

\section{Introduction}
\label{INTRODUCTION}

The Balitsky-Kovchegov (BK) equation \cite{Balitsky:1995ub,Kovchegov:1999yj,Kovchegov:1999ua} is a central tool for understanding the initial conditions in hadronic collisions in situations where transverse gluon density approaches unitarity limits.  At leading order in large $N_c$ and small $\alpha_s$, the equation reads
 \eqn{BKleading}{
  {\partial S(r_1,r_2;Y) \over \partial Y} = 
    {\bar\alpha_s \over 2\pi} \int d^2 z \, {|r_1-r_2|^2 \over |r_1-z|^2 |r_2-z|^2} \left[ 
      S(r_1,z;Y) S(z,r_2;Y) - S(r_1,r_2;Y) \right] \,,
 }
where 
 \eqn{alphabarDef}{
  \bar\alpha_s \equiv {\alpha_s N_c \over \pi}
 }
is essentially the 't Hooft coupling, and $r_1$, $r_2$, and $z$ are positions in the transverse plane, conveniently thought of as complex numbers.  The function
 \eqn{BKSdef}{
  S(r_1,r_2;Y) \equiv {1 \over N_c} \langle \tr U^\dagger(r_1) U(r_2) \rangle_Y
 }
is the correlator of fundamental Wilson lines in the presence of a hadronic target, evaluated at rapidity $Y$ relative to the target.  The range of $S$ is $[0,1]$, and for fixed $Y$ one usually requires $S(r_1,r_2;Y) \to 1$ as $r_1 \to r_2$, while $S(r_1,r_2;Y) \to 0$ as $r_1$ moves far away from $r_2$.  The quantity
 \eqn{NForm}{
  {\cal N}(r_1,r_2;Y) \equiv 1-S(r_1,r_2;Y)
 }
is the amplitude for a quark-antiquark dipole, characterized by position $r_1$ for the quark and $r_2$ for the anti-quark, to scatter off a hadronic target in a collision where the difference in rapidity between the dipole and the target is $Y$.  A common application of the BK equation (though far from the only one) is to study deep inelastic scattering (DIS), in particular the total cross-section of a virtual photon off of a proton.  In this application, the scattering is presumed to be dominated by the overlap of the photon with a quark-antiquark dipole, whose likelihood to interact with the hadronic target is then described by ${\cal N}(r_1,r_2;Y)$.

A standard simplification is to require that $S$ depends on $r_1$ and $r_2$ only through $|r_1-r_2|$.  Physically, this means that the dipole size is assumed to be much smaller than the typical size of the distribution of hadronic matter with which it collides.  This paper is focused on the observation that an equally valid simplification of the {\it leading order} equation \eno{BKleading} is to require that $S$ depends on $r_1$ and $r_2$ only through the distance function
 \eqn{ChordalDistance}{
  d_q(r_1,r_2) \equiv {|r_1-r_2| \over \sqrt{(1+q^2 |r_1|^2) (1+q^2 |r_2|^2)}} \,,
 }
where $q$ is an arbitrary parameter with dimensions of inverse length.  $1/q$ is essentially the size of the hadronic target.  The reason that solutions to \eno{BKleading} of the form 
 \eqn{ConformalForm}{
  S(r_1,r_2;Y) = S_q(d_q(r_1,r_2);Y)
 }
must exist is that the functional form on the right hand side is the one that respects an $SO(3)$ subgroup of the symmetry group $PSL(2,{\bf C})$ of the measure factor $d^2 z \, {|r_1-r_2|^2 \over |r_1-z|^2 |r_2-z|^2}$.  This is essentially the same $SO(3)$ subgroup as the one used in \cite{Gubser:2010ze,Gubser:2010ui} to find generalizations of Bjorken flow with finite transverse extent.  Because two out of the three generators of $SO(3)$ are conformal Killing vectors rather than true Killing vectors, solutions of the form \eno{ConformalForm} will be modified by the breaking of conformal invariance, which occurs at next-to-leading order (NLO) in the BK equation through the running of $\alpha_s$.  The ansatz \eno{ConformalForm} is similar in spirit to the work of \cite{Hatta:2009nd} on solutions to the Banfi-Marchesini-Smye equation preserving a $SO(2,1)$ subgroup of the conformal group.

The organization of the rest of this paper is as follows.  In section~\ref{SOLUTIONS} I briefly review some of the standard lore on solutions of the leading-order BK equation, including the key concepts of saturation scale and geometric scaling.  In section~\ref{CONFORMAL} I explain the $SO(3)$ symmetry more fully and show how this symmetry dictates the functional form of the distance function $d_q(x,y)$.  I will also demonstrate that the $SO(3)$ symmetry, together with some weak additional assumptions, dictates the following dependence of the saturation scale on the impact parameter $b$:
 \eqn{sigmaApprox}{
  Q_s(b;Y) = {Q_s^{\rm max}(Y) \over 1 + q^2 b^2} \,,
 }
provided $Q_s^{\rm max}(Y) \gg q$, which occurs at large enough rapidity.  In section~\ref{DIS}, I explain how $S(r_1,r_2;Y)$ is converted into a prediction of DIS cross-sections, and point out that although the form \eno{sigmaApprox} is satisfactory at large $Q$, it leads to a total cross-section $\sigma_{\gamma^*p}$ that increases somewhat too quickly as $Q$ becomes small.  This mismatch is due to the failure of conformally invariant solutions to capture the rapid falloff of energy density at sufficiently large transverse radius.  Conformal symmetry has also been used in \cite{Cornalba:2008sp}, along with some additional dynamical intuition from AdS/CFT, to study DIS cross-sections at small Bjorken $x$.

\section{Some well-known properties of the BK equation}
\label{SOLUTIONS}

Reviews of the large literature on the BK equation and related concepts include \cite{Mueller:2001fv,JalilianMarian:2005jf,Blaizot:2011pa}.  Earlier literature, for example \cite{Gribov:1984tu,Lipatov:1996ts}, shows that many aspects of the dynamics were understood before the BK equation was introduced in its modern form.  The remarks in this section are intended to provide non-specialists with some perspective on how the BK equation is used.

It is often convenient to replace $r_1$ and $r_2$ by the combinations
 \eqn{rAndb}{
  r = r_1 - r_2 \qquad\hbox{and}\qquad b = {r_1 + r_2 \over 2} \,.
 }
For consistency with the notation of \eno{BKleading}, $r$ and $b$ are most properly regarded as complex numbers.  However, when no ambiguity seems likely, I will abuse notation by allowing $r$ and $b$ to denote the magnitudes $|r|$ and $|b|$, respectively.  In particular, throughout this section I will restrict attention to solutions to the BK equation which depend only on $|r|$, and I will use $r$ to denote this positive real quantity.  That is,
 \eqn{Srestrict}{
  S(r_1,r_2;Y) = S(r;Y) \,.
 }

An important and well-studied class of solutions is those which tend toward a scaling form at large $Y$:
 \eqn{GeometricScaling}{
  S(r;Y) \to \hat{S}(\hat{r}) \qquad\hbox{as}\quad Y \to \infty \,,
 }
where the scaling variable is
 \eqn{tauDef}{
  \hat{r} = Q_s(Y) r
 }
and $Q_s(Y)$ grows exponentially with $Y$:
 \eqn{QsGrowth}{
  Q_s(Y) = Q_0 e^{{c \over 2} \bar\alpha_s Y} \qquad\hbox{where}\quad
   c \approx 4.88 \,.
 }
The estimate of $c$ quoted here is from \cite{Mueller:2002zm,Munier:2003sj}, where it is also argued that there are non-exponential corrections to \eno{QsGrowth}.  $Q_s(Y)$ is called the saturation scale.  Intuitively, $1/Q_s$ is the characteristic transverse size of gluons whose occupation numbers become comparable to unity.  A typical value for $Q_s$ in the description of deep inelastic scattering from HERA \cite{Derrick:1995ef,Aid:1996au,Adloff:1997mf,Breitweg:2000yn} is $1\,{\rm GeV}$ \cite{GolecBiernat:1998js}, though to assess precisely what this value means one must obviously have a specific form of $\hat{S}(\hat{r})$ in mind.  Dependence of $S(r;Y)$ and related quantities only on $\hat{r}$ (or mainly on $\hat{r}$) is known as geometric scaling \cite{GolecBiernat:1998js,Stasto:2000er} (see however the discussion following \eno{GeometricTau} for a refinement of what geometric scaling means when referring to DIS data).  Intuitively, geometric scaling says that the scattering amplitude at rapidity $Y$ for a dipole of transverse size $r$ depends only on the number of gluons at the relevant rapidity in the part of the target sampled by the dipole, which is proportional to $r^2 Q_s(Y)^2$.

It has been argued \cite{Munier:2004xu} that the approach \eno{GeometricScaling} to geometric scaling is a universal feature of all solutions $S(r;Y)$ to the BK equation satisfying certainly physically motivated conditions, in particular $|1-S(r;Y)| \ll r^{2\gamma_c}$ at some fixed $Y$ for small $r$, where\footnote{For the analysis of geometric scaling, it is convenient to start by defining
 \eqn{NkDef}{
  \hat{\cal N}(k;Y) = \int d^2 r \, e^{-i k \cdot r} {1 - S(r;Y) \over 2\pi r^2} \,,
 }
where $k$ and $r$ are regarded as vectors in ${\bf R}^2$.  Then, remarkably, \eno{BKleading} can be shown to be equivalent to
 \eqn{BKlog}{
  {\partial \hat{\cal N} \over \partial Y} = \bar\alpha_s \, \chi(-\partial_L)
    \hat{\cal N} - \bar\alpha_s^2 \, \hat{\cal N}^2 \,,
 }
where $L = \log k^2$ and $\chi(\gamma) = 2\psi(1) - \psi(\gamma) - \psi(1-\gamma)$.  Here $\psi(z) = d\log\Gamma(z)/dz$.  The exponent $\gamma_c$ in \eno{gammacValue} is defined as the solution to the equation $\gamma \chi'(\gamma) = \chi(\gamma)$.}
 \eqn{gammacValue}{
  \gamma_c \approx 0.628 \,.
 }
Qualitatively, the claim being made in \eno{GeometricScaling} is that hadronic collisions have universal behavior at high energies, independent of the precise nature of the colliding particles.

A certain amount is known about the scaling function $\hat{S}(\hat{r})$ that controls the asymptotic large $Y$ behavior:
 \begin{itemize}
  \item For $\hat{r} \ll 1$, one has $\hat{S}(\hat{r}) = 1 + C_1 \hat{r}^{2\gamma_c} \log\hat{r} + \ldots$, where $C_1>0$ is a constant \cite{Munier:2004xu}.
  \item For $\hat{r} \sim 1$, the numerical results of \cite{Albacete:2004gw} fit well to the form $\hat{S}(\hat{r}) = 1/(1+\hat{r}/\hat{r}_*)^2$, where $\hat{r}_*$ is a constant.
  \item For $\hat{r} \gg 1$, one has $\hat{S}(\hat{r}) = C_2 e^{-{4 \over c} \log^2 \hat{r}} + \ldots$, where $C_2>0$ is a constant and $c$ is the same constant as the one appearing in \eno{QsGrowth}.
 \end{itemize}
However, it is far from clear that large $Y$ asymptotics of the solutions of the leading-order BK equation are directly relevant for phenomenology.  Parton fraction $x \sim 10^{-4}$, which approximately characterizes the HERA deep inelastic scattering data as well as mid-rapidity LHC phenomena, corresponds to $Y = -\log x = 9.2$, whereas numerical investigations including \cite{Albacete:2004gw} show considerable evolution of scaling forms even for $Y > 20$.  So initial conditions at some modest value of $Y$ are important.  One favorite is the Golec-Biernat-W\"usthoff model \cite{GolecBiernat:1998js}
 \eqn{GBWinit}{
  S_{\rm GBW}(r;Y_0) = e^{-{1 \over 4} Q_{s0}^2 r^2} \,,
 }
where $Y_0$ and $Q_{s0}$ are parameters.

An additional challenge is that with reasonable values of the coupling, for example $\bar\alpha_s = 0.2$, the growth \eno{QsGrowth} is substantially more rapid than a comparison with DIS data supports.  This situation is substantially improved by the inclusion of next-to-leading-order effects: see for example \cite{Albacete:2009fh}.

\section{Exploiting conformal symmetry}
\label{CONFORMAL}

I want to consider distributions of hadronic matter which are non-uniform in the transverse plane.  In the literature on the BK equation, this is relatively unexplored territory.  In \cite{Mueller:2001fv} one can find a form for the dipole scattering amplitude based on treating a nucleus as a sphere of uniform density.  In \cite{Ferreiro:2002kv} one can find calculations related to the Froissart bound based on the assumption of a factorized form ${\cal N}(r_1,r_2) = {\cal N}(r) S(b)$.  Most relevant for our investigations, in \cite{Iancu:2007st} one can find the expression \eno{sigmaApprox} as an ansatz for the dependence of the saturation scale on impact parameter.  Earlier related work includes \cite{Munier:2001nr,Munier:2003bf,Kowalski:2003hm}, which consider how diffractive processes constrain the impact parameter dependence of dipole scattering amplitudes; and \cite{GolecBiernat:2003ym}, in which solutions to the BK equation with explicit impact parameter dependence were investigated numerically.

In general, it's hard work to study the BK equation when $S$ has some general dependence on complex $r_1$ and $r_2$ rather than dependence only on $r = |r_1-r_2|$, simply because the general form $S(r_1,r_2;Y)$ has five real independent variables, as compared to two independent real variables in $S(r;Y)$.  Dependence of $S$ only on $r=|r_1-r_2|$ and $Y$ amounts to the imposition of $ISO(2)$ symmetry.  The action of $ISO(2)$ on a complex number $z$ (describing, as usual, a position in the transverse plane) is
 \eqn{ISOtwo}{
  z \to \alpha z + \beta
 }
where $|\alpha|=1$ and $\beta$ is any complex number.  Imposing $ISO(2)$ symmetry on a function $S(r_1,r_2;Y)$ means that we demand that $S(r_1,r_2;Y)$ is invariant if we send $r_1 \to \alpha r_1 + \beta$ and $r_2 \to \alpha r_2 + \beta$ at the same time (and leave $Y$ unchanged).

In \cite{Gubser:2010ze} I showed in the context of relativistic hydrodynamics that an interesting solution-generating technique was to replace invariance under $ISO(2)$ by invariance under a group $SO(3)$ of conformal isometries of ${\bf R}^{3,1}$.  I will exploit the same strategy here.  The reason it works is that the measure and kernel of the leading-order BK equation is invariant under the group $PSL(2,{\bf C})$ of linear fractional transformations (LFTs),\footnote{$SL(2,{\bf C})$ is the group of matrices $\begin{pmatrix} \alpha & \beta \\ \gamma & \delta \end{pmatrix}$ where $\alpha$, $\beta$, $\gamma$, and $\delta$ are complex with $\alpha\delta-\beta\gamma=1$.  $PSL(2,{\bf C})$ is $SL(2,{\bf C})$ quotiented by the ${\bf Z}_2$ which simultaneously flips the sign of $\alpha$, $\beta$, $\gamma$, and $\delta$.  LFTs of the form \eno{SLtwoC} are in unique correspondence with elements of $PSL(2,{\bf C})$.} which act on $z$ by
 \eqn{SLtwoC}{
  z \to {\alpha z + \beta \over \gamma z + \delta} \,,
 }
where $\alpha$, $\beta$, $\gamma$, and $\delta$ are complex.  A simple way to check this invariance is to consider only the holomorphic half of the measure and kernel, namely the one-form
 \eqn{omegaDef}{
  \omega = {r_1-r_2 \over (r_1-z)(r_2-z)} dz \,.
 }
The claim is that $\omega$ is invariant under all linear fractional transformations, where we understand that the LFT is supposed to act on $r_1$, $r_2$, and $z$.  To check this claim, first note that it is obvious for linear transformations \eno{ISOtwo} (even when $\alpha$ is not unimodular), and that the full group $PSL(2,{\bf C})$ is generated by linear transformations together with the inversion $z \to -1/z$.  So the only thing to check is that $\omega$ is invariant under $z \to -1/z$ (with $r_1 \to -1/r_1$ and $r_2 \to -1/r_2$ at the same time).  This is a straightforward exercise.  Conformal invariance of the BFKL kernel has been thoroughly appreciated in the literature (see for example \cite{Lipatov:1996ts}), and the invariance of the BK equation that I have just described has been noted explicitly, for example in \cite{GolecBiernat:2003ym}.

Just as $ISO(2)$-invariance requires that $S$ should depend on $r_1$ and $r_2$ only through the combination $|r_1-r_2|$, so one would expect invariance under some other subgroup $G \in PSL(2,{\bf C})$ to require that $S$ should depend on $r_1$ and $r_2$ only through some other simple combination.  If such a symmetry is imposed on $S(r_1,r_2;Y)$ at some fixed $Y$, then the $PSL(2,{\bf C})$ invariance of the leading-order BK equation ensures that it persists at all $Y$.  The obvious example to start with is an $SO(3)$ subgroup which includes rotations around the beam axis (that is, rotations of the complex plane preserving the origin).  There is a one-parameter family of such subgroups, characterized by a length scale which I will denote $1/q$, following \cite{Gubser:2010ze}.  Besides rotations preserving $z=0$, $SO(3)_q$ includes the LFTs
 \eqn{NextLFT}{
  qz \to {qz\cos{\theta \over 2} + \sin{\theta \over 2} \over 
    -qz\sin{\theta \over 2} + \cos{\theta \over 2}} \,,
 }
which preserve the points $z=\pm i/q$.  The parameter $\theta \in [0,2\pi)$ uniquely labels these conformal isometries of the complex plane.  The rest of $SO(3)_q$ can be obtained by composing rotations preserving $z=0$ with transformations of the form \eno{NextLFT}.  It is useful to recall that there is a conformal map, the so-called stereographic projection, which maps the complex plane to the unit sphere $S^2$.  Let's parametrize the $S^2$ by polar angles $(\theta,\phi)$ such that $\theta$ runs from $0$ at the north pole to $\pi$ at the south pole, and $\phi$ runs from $0$ to $2\pi$ and labels longitude.  Then the stereographic map is
 \eqn{Stereographic}{
  |qz| = \tan {\theta \over 2} \qquad\qquad \arg z = \phi \,,
 }
and the transformations \eno{NextLFT} can be understood as ordinary rotations of $S^2$ which preserve two opposite points located on the equator.

The $SO(3)_q$ which I have described is essentially the same as the one in \cite{Gubser:2010ze}, and its original motivation came from collisions of gravitational shocks in $AdS_5$ \cite{Gubser:2008pc,Gubser:2009sx}.\footnote{Interestingly, there is a literature on comparing AdS/CFT calculations involving gravitational shocks to DIS data: see for example \cite{Albacete:2008ze}.  This work is, however, largely orthogonal to the present work, because the focus here is on the use of $SO(3)_q$ as an organizing principle for dependence of scattering amplitudes on impact parameter, whereas the focus in \cite{Albacete:2008ze} is on the AdS/CFT prescription for computing Wilson loops.}   More precisely, each element of $SO(3)_q$ is a conformal isometry of the complex plane which can be extended to a conformal isometry of ${\bf R}^{3,1}$, and this extended isometry is an element of the group $SO(3)_q$ picked out in \cite{Gubser:2010ze}.  This is most easily seen at the level of derivative operators representing the generators of $SO(3)_q$.  The infinitesimal generator of transformations \eno{NextLFT} is
 \eqn{xiDef}{
  \zeta_\perp \equiv (1+q^2 z^2) \partial_z + (1+q^2 \bar{z}^2) \partial_{\bar{z}} \,.
 }
If we write $z = x^1 + i x^2$, then
 \eqn{zetaReal}{
  \zeta_\perp = 
   {\partial \over \partial x^1} + q^2 \left[ 2x^1 \left( x^1 {\partial \over \partial x^1} + 
    x^2 {\partial \over \partial x^2} \right) - 
    \left( (x^1)^2 + (x^2)^2 \right) {\partial \over \partial x^1} \right] \,.
 }
Now recall the form of the symmetry generator used in \cite{Gubser:2010ze} to define $SO(3)_q$:
 \eqn{zetaForm}{
  \zeta = {\partial \over \partial x^1} + q^2 \left[ 2 x^1 x^\mu {\partial \over \partial x^\mu} - 
    x^\mu x_\mu {\partial \over \partial x^1} \right] \,,
 }
where $x^\mu = (t,x^1,x^2,x^3)$ and $x_\mu = (-t,x^1,x^2,x^3)$.  Evidently, $\zeta_\perp = \zeta$ provided $t^2 = (x^3)^2$.  In other words, $\zeta_\perp$ on the light-front locus $t^2 = (x^3)^2$ extends to $\zeta$ on all of ${\bf R}^{3,1}$.

Having established the main properties of the $SO(3)_q$ symmetry, let's ask next what real function of complex $r_1$ and $r_2$ is invariant under it.  That there should be essentially one such function makes sense from the point of view of counting real parameters: $r_1$ and $r_2$ together comprise four real parameters, and each of the three generators of $SO(3)_q$ can be thought of as constraining one.  From a more geometrical point of view, it's obvious that there should be essentially only one $SO(3)_q$-invariant interval, because on the unit sphere we know what it is: geodesic distance between the two points in question.  More convenient than geodesic distance for many purposes is chordal distance: if we embed $S^2$ into ${\bf R}^3$ in the standard way (i.e.~as the unit sphere centered on the origin), then the chordal distance between two points on $S^2$ is the length of the line segment connecting them in ${\bf R}^3$.  Using the definition \eno{Stereographic} of the stereographic map, one may straightforwardly check that the chordal distance between the stereographic images of two points $r_1$ and $r_2$ is $2q d_q(r_1,r_2)$, where $d_q(r_1,r_2)$ is the distance function defined in \eno{ChordalDistance}.  It is possible to show directly that $d_q(r_1,r_2)$ is unchanged when $r_1$ and $r_2$ are replaced by $SO(3)_q$ images of themselves, for instance under the LFT map \eno{NextLFT}.

The discussion so far makes clear that there are solutions of the BK equation that respect $SO(3)_q$ invariance, and that they take the form
 \eqn{ConformalFormAgain}{
  S(r_1,r_2;Y) = S_q(d_q(r_1,r_2);Y) \,.
 }
It remains to calculate $S_q(d_q;Y)$.  This is hard, in the sense that no analytical results are likely to be available.  However, the physically interesting regime to consider is $Q_s \gg q$: that is, the saturation scale is much smaller than the overall size of the target.  Then there is a simple way of generating approximate solutions: If $S(r;Y)$ is an $ISO(2)$ symmetric solution which vanishes quickly away from small $r = |r_1-r_2|$, then one need only replace $r \to d_q(r_1,r_2)$ in order to obtain an $SO(3)$-invariant function which is an approximate solution to \eno{BKleading}.  The reason this works is that the $SO(3)$-invariant function is significantly positive only when $q d_q(r_1,r_2) \ll 1$.  So it doesn't ``notice'' the curvature of $S^2$, and acts the way solutions do on ${\bf R}^2$.  As a special case, we can start from an $ISO(2)$-invariant scaling solution $S(r;Y) = \hat{S}(\hat{r})$.  The corresponding approximate $SO(3)$-invariant solution is
 \eqn{Sscaling}{
  S(r_1,r_2;Y) = \hat{S}(Q_s^{\rm max}(Y) d_q(r_1,r_2)) \,.
 }
In the $ISO(2)$-invariant solution, the quantity $Q_s^{\rm max}(Y)$ is exactly the $Q_s(Y)$ that came up in \eno{tauDef} and \eno{QsGrowth}; but I have renamed it in \eno{Sscaling} in order to make clear its new meaning in the $SO(3)$-invariant solution as the maximum of the saturation scale over the transverse plane.

When $q d_q(r_1,r_2) \ll 1$, one can make the further approximation
 \eqn{dqApprox}{
  d_q(r_1,r_2) \approx {|r_1-r_2| \over 1 + q^2 \left| {r_1+r_2 \over 2} \right|^2} 
    = {|r| \over 1 + q^2 |b|^2} \,.
 }
Plugging \eno{dqApprox} into \eno{Sscaling}, one finds
 \eqn{CSagain}{
  S(r_1,r_2;Y) \approx \hat{S}\left(Q_s(b;Y) r \right)
 }
where
 \eqn{sigmaApproxAgain}{
  Q_s(z;Y) = {Q_s^{\rm max}(Y) \over 1 + q^2 b^2}
 }
is the saturation scale at a position $z$ in the transverse plane.  The form \eno{sigmaApprox} has been considered previously \cite{Iancu:2007st} (see also \cite{Bondarenko:2003ym}), but now we can see precisely what its distinguishing feature is: It is the $SO(3)_q$-invariant choice for the dependence of the saturation scale on transverse position.

To derive \eno{sigmaApproxAgain} I used the assumption of geometric scaling.  However, much weaker assumptions suffice.  Assume we have $SO(3)$-invariance, so that \eno{ConformalFormAgain} holds for some function $S_q(d_q;Y)$.  Assume further that for fixed $d_q>0$, one has $S_q(d_q;Y) \to 0$ as $Y \to \infty$.  Lastly, assume that for any fixed value of $Y$, $S_q(d_q;Y)$ is a monotonically decreasing function of $d_q$.  Let's {\it define} $Q_s^{\rm max}(Y)$ through the equation
 \eqn{QsMaxDef}{
  S_q\left( {1 \over Q_s^{\rm max}(Y)}; Y \right) = \kappa \,,
 }
where $\kappa \in (0,1)$ is some pre-specified number.  $Q_s^{\rm max}(Y)$ is unique because of the monotonicity property.  $Q_s^{\rm max}(Y)$ exists for sufficiently large $Y$ because $S_q \to 0$ as $Y \to \infty$ for $d_q > 0$.\footnote{This is a non-trivial observation because $S_q$ does {\it not} in general reach arbitrarily small values.  If we specify a value of $\kappa$ that is not particularly small, say $\kappa = 1/2$, then $Q_s^{\rm max}(Y)$ as defined in \eno{QsMaxDef} exists for all $Y$ above a modest threshold.}  $Q_s^{\rm max}(Y) \to \infty$ as $Y \to \infty$, again because $S_q \to 0$ as $Y \to \infty$ away from $d_q=0$.  The regime we are interested in is $Q_s^{\rm max}(Y) \gg q$.  In this regime, up to ambiguities suppressed by powers of $q/Q_s^{\rm max}(Y)$, we can define the saturation scale $Q_s(b;Y)$ as the solution the equation
 \eqn{QsConformalDef}{
  d_q\left( b - {1 \over 2 Q_s(b;Y)}, b + {1 \over 2 Q_s(b;Y)} \right) = 
    {1 \over Q_s^{\rm max}(Y)} \,.
 }
Using the approximation \eno{dqApprox} to simplify the left-hand side of \eno{QsConformalDef}, we arrive immediately at \eno{sigmaApprox}.

\section{Deep inelastic scattering}
\label{DIS}

I will first briefly review how the amplitude $S(r_1,r_2;Y)$ is used to calculate certain cross-sections in deep inelastic scattering.  For a more complete account, see for example \cite{GolecBiernat:1998js,Mueller:2001fv}.  A key quantity for comparison with experiment is the total cross-section $\sigma_{\gamma^*p}$ of a virtual photon, denoted $\gamma^*$, to scatter off the hadronic target, which at HERA is a proton.  Standard calculations, summarized for example in \cite{JalilianMarian:2005jf}, lead to the formulas
 \eqn{sigmaSum}{
  \sigma_{\gamma^*p}(x,Q^2) = \sigma_T(x,Q^2) + \sigma_L(x,Q^2)
 }
where
 \eqn{sigmaTL}{
  \sigma_{T,L}(x,Q^2) = \int d^2 r_1 \int d^2 r_2 \int_0^1 dv \,
    |\Psi_{T,L}(r,v,Q^2)|^2 {\cal N}(r_1,r_2;Y) \,,
 }
and I have again used the convenient notation $r = |r_1-r_2|$.  Also,
 \eqn{PsiTL}{
  |\Psi_T|^2 &= {3\alpha_{\rm em} \over \pi^2} \sum_f
     e_f^2 \left( \left[ v^2 + (1-v)^2 \right] \bar{Q}_f^2 K_1(\bar{Q}_f r)^2 + 
       m_f^2 K_0(\bar{Q}_f r)^2 \right)  \cr
  |\Psi_L|^2 &= {3\alpha_{\rm em} \over \pi^2} \sum_f
     e_f^2 \left( 4 Q^2 v^2 (1-v)^2 K_0(\bar{Q}_f r)^2 \right)
 }
where $K_0$ and $K_1$ are modified Bessel functions.  Also,
 \eqn{QfDef}{
  \bar{Q}_f^2 = v(1-v) Q^2 + m_f^2 \,.
 }
$\Psi_T$ and $\Psi_L$ are the wave functions for splitting the transverse and longitudinal polarizations of the virtual photon into a quark at transverse position $r_1$ and an anti-quark at transverse position $r_2$, with $r = |r_1-r_2|$ as usual.  $v$ is the momentum fraction of the photon carried by the quark.  The subscript $f$ in the sums indicates the quark species, and $m_f$ and $e_f$ are the mass and charge of each quark.  The dependence of $|\Psi_{T,L}|^2$ on $r_1$ and $r_2$ only through $r$ amounts to the assumption that the photon is equally likely to be anywhere in the transverse plane.  $Q^2$ is the photon's virtuality (positive for a spacelike photon), and $Y = \log {1 \over x}$ where $x$ is Bjorken's variable.  

A well-known analysis \cite{GolecBiernat:1998js,Stasto:2000er} adopts the following simplified form of \eno{sigmaTL}:
 \eqn{sigmaGBW}{
  \sigma_{T,L}(x,Q^2) = \sigma_0 \int d^2 r \int_0^1 dv \, 
    |\Psi_{T,L}(r,v,Q^2)|^2 {\cal N}_{\rm GBW}(r;Y) \,,
 }
where $\sigma_0$ is a constant with units of area and
 \eqn{NGBW}{
  {\cal N}_{\rm GBW}(r;Y) = 1 - e^{-{1 \over 4} Q_s(Y)^2 r^2} \,.
 }
In the limit $m_f \to 0$, the geometric scaling property of \eno{NGBW} translates into a dependence of $\sigma_{T,L}(x,Q^2)$ on $Y$ and $Q^2$ only through the dimensionless ratio
 \eqn{GeometricTau}{
  \tau \equiv {Q^2 \over Q_s(Y)^2} \,.
 }
Geometric scaling in DIS data refers, most properly, to the dependence of $\sigma_{\gamma^*p}(x,Q^2)$ and related quantities only on $\tau$.  Evidently, this can only work if $Q_s(Y)$ is chosen properly, and a good fit to the data is achieved using
 \eqn{QsDepend}{
  Q_s(Y) = (1\,{\rm GeV}) e^{0.14 (Y-8.1)}
 }
and $\sigma_0 = 23\,\mu{\rm b}$, with a sum over the three flavors $u$, $d$, $s$ with $m_f = 140\,{\rm MeV}$ (a value small enough so that it only slightly modifies geometric scaling) \cite{GolecBiernat:1998js}.  The simplification \eno{sigmaGBW} leaves something to be desired, because instead of integrating a scattering amplitude ${\cal N}(r_1,r_2;Y)$ over the positions of both the quark and the anti-quark, one starts by assuming translational invariance ${\cal N}(r_1,r_2;Y) = {\cal N}(r;Y)$.  Translational invariance of the hadronic target automatically makes all cross-sections infinite, and this is avoided in \eno{sigmaGBW} by replacing the integration over impact parameter by the factor $\sigma_0$.

To understand the main implications of $SO(3)$-invariant scattering amplitudes \eno{Sscaling} for DIS, let's consider the following ansatz:
 \eqn{NxyAnsatz}{
  {\cal N}(r_1,r_2;Y) = 1 - e^{-{1 \over 4} Q_s(b;Y)^2 r^2} \,,
 }
where $r = |r_1-r_2|$ and $b = \left| {r_1 + r_2 \over 2} \right|$.  I plan to use the expression \eno{sigmaApprox} for $Q_s(b;Y)$, but to start with let's consider a general dependence of $Q_s$ on the magnitude of $b$ and on $Y$.  A useful simplification is to note that when $m_f=0$,
 \eqn{PsiSimplify}{
  \int_0^1 dv \, \left( |\Psi_T(r,v,Q^2)|^2 + |\Psi_L(r,v,Q^2)|^2 \right) \approx 
    {\alpha_{\rm em} Q^2 \over 3\pi^2} {1 \over (Qr)^2 + (Qr)^4/4} \,.
 }
This form, though approximate, is convenient because it allows all but one integral in the expression for $\sigma_{\gamma^*p}$ to be performed explicitly: starting from \eno{sigmaSum}-\eno{PsiTL} and \eno{NxyAnsatz}, one finds
 \eqn{FoundSigma}{
  \sigma_{\gamma^*p} &= {8\alpha_{\rm em} \over 3} Q^2 \int_0^\infty b \, db
    \int_0^\infty r \, dr \, {1 - e^{-{1 \over 4} Q_s(b;Y)^2 r^2} \over 
      (Qr)^2 + (Qr)^4/4}  
    = {4\alpha_{\rm em} \over 3} \int_0^\infty b \, db \, G\left( {Q^2 \over Q_s(b;Y)^2}
       \right)
 }
where
\def\Ei{\mop{Ei}}
 \eqn{Gexplicit}{
  G(\theta) = \gamma - \log\theta - e^{1/\theta} \Re\Ei(-1/\theta) \,.
 }
Here $\Ei$ is the exponential integral function, and $\gamma$ is Euler's constant.  Let's define
 \eqn{tauAgain}{
  \tau = {Q^2 \over \overline{Q}_s(Y)^2} \qquad\hbox{and}\qquad
   u(b;Y) = {Q_s(b;Y) \over \overline{Q}_s(Y)} \,,
 }
so that \eno{FoundSigma} may be re-expressed as
 \eqn{sigmaAgain}{
  \sigma_{\gamma^* p} = {4\alpha_{\rm em} \over 3} \int_0^\infty b \, db \, 
    G\left( {\tau \over u(b;Y)^2} \right) \,.
 }
In \eno{tauAgain}, $\overline{Q}_s(Y)$ is an average or characteristic value of $Q_s$, whose precise definition we do not need to specify at this stage.  In order for $\sigma_{\gamma^* p}$ to depend on $Q^2$ and $Y$ only through the combination $\tau$ defined in \eno{tauAgain}, $u$ must be a function only of $b$, not of $Y$.  For $SO(3)$-invariant solutions, according to \eno{sigmaApprox}, we have
 \eqn{uExample}{
  u = {Q_s^{\rm max} / \overline{Q}_s \over 1+q^2 b^2} \,,
 }
so $u$ is independent of $Y$ precisely if $\overline{Q}_s$ is a fixed fraction of $Q_s^{\rm max}$. 

It is easy to see that $G(\theta)$ is a monotonically decreasing function with
 \eqn{GthetaLimits}{
  G(\theta) = \left\{ \seqalign{\span\TL &\qquad \span\TT}{
    -\log\theta + \ldots & for $\theta \ll 1$  \cr
    {1 \over \theta} \log \theta + \ldots & for $\theta \gg 1$.} \right.
 }
where $\ldots$ indicates subleading terms.  As a rough estimate, in the expansions \eno{GthetaLimits}, we may in replace $\log\theta$ by $-1$ for $\theta \ll 1$ and $+1$ for $\theta \gg 1$, and split up the last integral in \eno{FoundSigma} as
 \eqn{SigmaRough}{
  \sigma_{\gamma^*p}(x,Q^2) \approx {4\alpha_{\rm em} \over 3} 
    \left[ \int_0^{b_*} b \, db + 
      \int_{b_*}^\infty b \, db \, {Q_s(b;Y)^2 \over Q^2} \right]
 }
where $b_*$ is determined implicitly by the equation
 \eqn{DetermineBstar}{
  Q_s(b_*;Y) = Q \,.
 }
If $Q$ is large, then \eno{DetermineBstar} has no solutions, and one must effectively set $b_*=0$ in \eno{SigmaRough}.  Then one finds $\sigma_{\gamma^*p} \propto 1/\tau$ where $\tau$ is defined as in \eno{tauAgain}, up to logarithmic corrections due to the coarseness of the approximations in \eno{SigmaRough}.  If $Q$ is small, then $b_*$ is large, and the first term in square brackets in \eno{SigmaRough} dominates.  In this situation, $\sigma_{\gamma^*p} \propto b_*^2$, again with logarithmic corrections.  In summary,
 \eqn{SigmaSummary}{
  \sigma_{\gamma^* p} \propto \left\{ \seqalign{\span\TL &\qquad \span\TT}{
    {1 \over \tau} & for $\tau \gg 1$  \cr
    b_*^2 & for $\tau \ll 1$ \,.} \right.
 }

The analysis so far has been for a general dependence of $Q_s$ on impact parameter $b$, provided the ratio $u$ defined in \eno{tauAgain} depends only on $b$, not $Y$.  Evidently, the large $\tau$ behavior $\sigma_{\gamma^*p} \propto 1/\tau$ is universal, and, pleasingly, this behavior is a good fit to the data \cite{GolecBiernat:1998js,Stasto:2000er}.  According to \eno{DetermineBstar} and \eno{SigmaSummary}, the small $\tau$ behavior of $\sigma_{\gamma^*p}$ is a probe of the behavior of $Q_s(b;Y)$ at large $b$---assuming, as seems reasonable, that $Q_s(b;Y)$ decreases monotonically to $0$ (or to values below $\Lambda_{\rm QCD}$) as $b$ increases.  If we assume the form \eno{sigmaApprox} for $Q_s(b;Y)$, then \eno{DetermineBstar} can be solved to give $\sigma_{\gamma^*p} \propto 1/\sqrt\tau$ for small $\tau$.  This is a faster increase with decreasing $\tau$ than is supported by the data: Referring to Figure~2 of \cite{Stasto:2000er}, we see that as $\tau$ decreases, $\sqrt\tau \sigma_{\gamma^*p}$ saturates to a maximum value near $\tau = 1$ and then decreases roughly like a small positive power of $\tau$ to as far as the data reaches, roughly $\tau = 2 \times 10^{-3}$ (see also figure~\ref{DIScomparison} below).  This decrease is consistent with a dependence $\sigma_{\gamma^*p} \propto 1/\sqrt[4]\tau$, though the width of the available region of $\tau$ is narrow enough that a firm conclusion about the functional form that describes it should not be drawn.

We should not be too surprised that conformal invariance predicts a faster increase in $\sigma_{\gamma^*p}$ at small $\tau$ than data supports: this increase is due to the relatively slow decrease of $Q_s(b;Y)$ at large $b$ exhibited by \eno{sigmaApprox}.  Confinement indicates that $Q_s$ should decrease more rapidly than $1/b^2$ once one reaches a sufficiently large value of $b$.

A weakness of the analysis I have presented is that the form \eno{NxyAnsatz} with $Q_s(b;Y)$ given by \eno{sigmaApprox} is only an approximate presentation of $SO(3)$-invariance, valid when $r \ll 1/q$.  A more precise expression of $SO(3)$-invariance would be to use \eno{Sscaling}; however this makes the integrations quite a bit more complicated.  There is a way of seeing that this weakness is not a serious problem, as follows.  The regime of parameters corresponding to HERA data is $Q$ between $0.2\,{\rm GeV}$ and $20\,{\rm GeV}$, with $Q_s$ on the order of $1\,{\rm GeV}$.  On the other hand, $q$ should be chosen to be approximately the inverse radius of the proton, i.e.~$0.2\,{\rm GeV}$.  Thus $Q \gsim q$, and we can expect the main contribution to the integrals defining $\sigma_{\gamma^*p}$ to come from $r \lsim 1/q$.  In short, in the physically interesting regime, we are seldom far from satisfying the desired inequality $r \ll 1/q$, so the use of \eno{sigmaApprox} is sufficient in order to understand the qualitative features of the dependence of $\sigma_{\gamma^*p}$ on $\tau$.

Another issue is that \eno{NxyAnsatz} does not solve the BK equation.  Clearly, an improved analysis would be desirable.  However, I expect that the essential feature of excessive growth of $\sigma_{\gamma^*p}$ at small $\tau$ is a durable consequence of conformal symmetry, since it is related to the absence of confinement.

Cutting off the integration over $b$ in \eno{FoundSigma} at some maximum value $b_{\rm max}$ approximately equal to the size of the proton improves the fit to measurements of the total cross-section $\sigma_{\gamma^* p}$, including the decrease in $\sqrt{\tau}\sigma_{\gamma^*p}$ as $\tau$ decreases well below $1$.  This is not too surprising since the cutoff in $b$ is a crude implementation of the effects of confinement.  For comparison to data, I used
 \eqn{sigmaBetter}{
  \sigma_{\gamma^* p} = {4\alpha_{\rm em} \over 3} \int_0^{b_{\rm max}} b \, db \, 
    \tilde{G}\left( {\tau \over u(b)^2} \right)
 }
with $u(b)$ given as in \eno{uExample} and $\tilde{G}(\theta)$ derived as in \eno{FoundSigma} from an improved approximation to the integrated photon wave function, namely
 \eqn{PsiBetter}{
  \int_0^1 dv \, \left( |\Psi_T(r,v,Q^2)|^2 + |\Psi_L(r,v,Q^2)|^2 \right) \approx 
    {\alpha_{\rm em} Q^2 \over 3\pi^2} {1 \over s^2 + s^4/4} \left( 1 + {s^2/2 \over
      (1+s^2/4)^2} \right)
 }
where $s=Qr$.  Calculations were performed with three massless flavors and $\alpha_{\rm em} = 1/133$.  It turns out that a good fit for $\sigma_{\gamma^* p}$ can be achieved over a significant range of choices for $q$ with a cutoff $b_{\rm max} = 0.69\,{\rm fm},$\footnote{It is reassuring that the value $b_{\rm max} = 0.69\,{\rm fm}$ is close to the root-mean-square charge radius of the proton, $R_p = 0.88\,{\rm fm}$, as measured in elastic processes, though perhaps it is a bit surprising to find $b_{\rm max} < R_p$.  Note however that $b_{\rm max}$ is sensitive to the overall normalization of $\sigma_{\gamma^*p}$, so it is affected by any uncertainties in the overall normalizations of the photon wave-function and of the cross-section.} provided that one allows $Q_s^{\rm max}/\overline{Q}_s$ to be adjusted as a fit parameter for each value of $q$.  See figure~\ref{DIScomparison}.
 \begin{figure}
  \hskip-0.55in\includegraphics[width=7.5in]{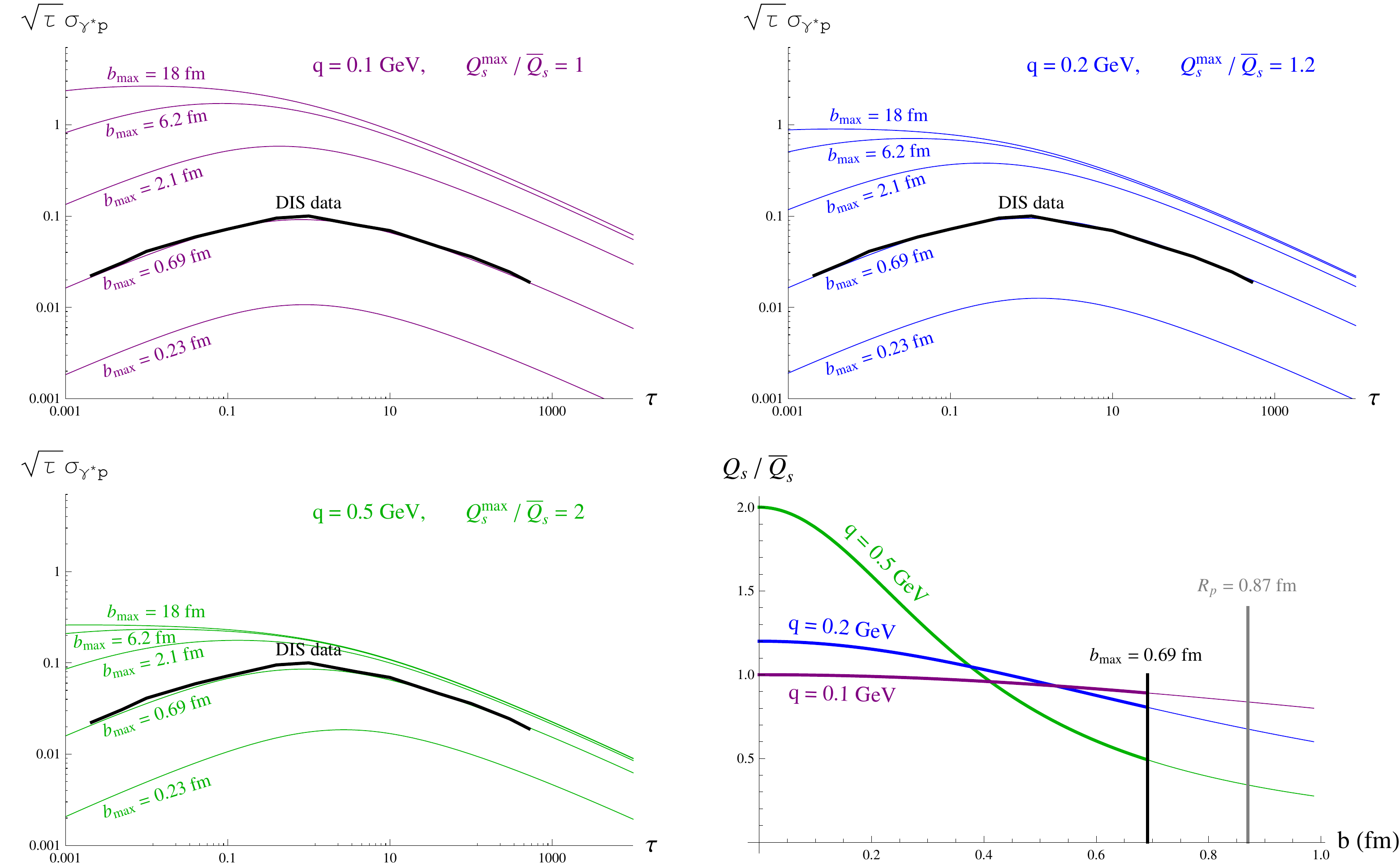}
  \caption{The plots of $\sqrt\tau \sigma_{\gamma^*p}$ show comparisons between predictions derived from \eno{sigmaAgain} for various values of $b_{\rm max}$ and DIS data, taken from Figure~2 of \cite{Stasto:2000er}.  In each plot of $\sqrt\tau \sigma_{\gamma^* p}$, $Q_s^{\rm max}/\overline{Q}_s$ was set to a constant that led to a good fit with data for $b_{\rm max} = 0.69\,{\rm fm}$.  The plot at lower right shows the dependence of $Q_s / \overline{Q}_s$ on $b$, given by \eno{uExample}, with the values of $q$ and $Q_s^{\rm max}/\overline{Q}_s$ as shown in the other three plots.}\label{DIScomparison}
 \end{figure}

It is well understood that the comparison of saturation ideas to total cross section $\sigma_{\gamma^*p}$ in DIS data can be made to work with a saturation scale $Q_s$ that is independent of impact parameter $b$ out to a hard cutoff beyond which it vanishes.  The plot with $q=0.1\,{\rm GeV}$ in figure~\ref{DIScomparison} is essentially a confirmation of this point.  But, as is evident from the other two values of $q$ shown, the data for $\sigma_{\gamma^*p}$ is also consistent with significant variation of $Q_s$, at least for the $SO(3)$-invariant functional form \eno{sigmaApprox} supplemented by the hard cutoff in $b$.  Admittedly, using $q=0.5\,{\rm GeV}$ pushes past the envelope where use of \eno{dqApprox} and \eno{CSagain} is uniformly reliable across the experimentally accessible range of $Q$.  Thus an improved analysis would again be desirable.  Comparison to diffractive cross-sections would also be helpful.

\section{Conclusions}
\label{CONCLUSIONS}

The main claims of this paper are contained in equations \eno{ConformalForm} and \eno{sigmaApprox}.  The first result is the statement that there are solutions to the leading order BK equation which respect an $SO(3)$ subgroup of the conformal group.  The $SO(3)$ subgroup is characterized by a parameter $q$ with dimensions of inverse length.  $SO(3)_q$-invariant solutions are almost as simple as the more widely studied translationally invariant solutions, where the scattering amplitude $S$ depends only on the dipole size, not its position in the transverse plane; indeed, translationally invariant solutions can be recovered from the $q \to 0$ limit of $SO(3)_q$-invariant solutions.  After mapping the transverse plane stereographically to a sphere, one can understand $SO(3)_q$-invariance as uniformity on the sphere.  Indeed, the distance function $d_q$ appearing in \eno{ChordalDistance} and \eno{ConformalForm} is the the chordal distance between the two images points on the sphere, which is obviously $SO(3)_q$-invariant.  The form of $d_q$ leads directly to the specific dependence of the saturation scale $Q_s$ on impact parameter shown in \eno{sigmaApprox}.

It is noteworthy that in solutions to the BK equation of the form \eno{ConformalForm}, the saturation scale doesn't change its functional dependence on impact parameter as the rapidity increases.  Heuristically, the hadronic target doesn't spread out at all, it just gets blacker all over.

Conformal invariance in QCD is modified by the running of the coupling, and most spectacularly by confinement.  So we should expect to start to have trouble matching data as the characteristic momentum scale approaches $\Lambda_{\rm QCD}$.  Indeed the analysis of section~\ref{DIS} showed too fast an increase of the cross-section $\sigma_{\gamma^*p}$ with $\tau = Q^2/Q_s^2$ as $\tau$ becomes small.  This is because conformal invariance predicts a fatter tail of energy density and saturation scale at large impact parameter than a confining theory like QCD will support.  However, this is hardly sufficient reason to abandon the conformal approach: when $Q_s^{\rm max} \gg \Lambda_{\rm QCD}$ there may be a substantial region in the transverse plane where the results \eno{ConformalForm} and \eno{sigmaApprox} provide a good leading order description of the variation of parton distribution functions over the transverse plane.  Using conformal invariance as an organizing principle for the early stages of a collision is likely to be especially useful when the early time dynamics itself is conformally invariant.  This is true in the glasma description \cite{Lappi:2006fp} (see also the earlier work \cite{Fries:2005yc}, and \cite{Fries:2006pv}), where the relevant dynamics is classical Yang-Mills theory.  To the extent that conformal symmetry is a good symmetry of both the initial state and the subsequent early-time dynamics, the analytic solution of \cite{Gubser:2010ze} should describe the fluid flow until such time as the non-conformal nature of the equation of state becomes important.

Clearly it would be desirable from a phenomenological standpoint to study deviations from $SO(3)_q$ invariance in solutions to the BK equation, as I did with A.~Yarom for hydrodynamics in \cite{Gubser:2010ui}.  For example, I would like to know whether and how fast the $SO(3)_q$-symmetric configuration is approached according to the leading-order dynamics \eno{BKleading}.  More ambitiously, one could inquire how NLO corrections cause the inverse size parameter $q$ to evolve with rapidity, and how deviations from conformality evolve according to NLO dynamics.  Best of all would be to develop a quantitative understanding of how deviations from conformality in the initial state propagate through to the hydrodynamical regime.

\section*{Acknowledgments}

I thank S.~Pufu, G.~Salam, L.-T.~Wang, and A.~Yarom for discussions.  This work was supported in part by the Department of Energy under Grant No.~DE-FG02-91ER40671, and by the Guggenheim Foundation.

\bibliographystyle{ssg}
\bibliography{bk}

\end{document}